\documentstyle[12pt]{article}
\include{epsf}
 
\newlength{\extralineskip}
\newcommand{\beq}{\begin{equation}}
\newcommand{\eeq}{\end{equation}}
\newcommand{\bd}{\begin{displaymath}}
\newcommand{\ed}{\end{displaymath}}
\def\bea{\begin{eqnarray}}
\def\eea{\end{eqnarray}}

\def\ba{\beq\new\begin{array}{c}}
\def\ea{\end{array}\eeq}

\def\inbar{\,\vrule height1.5ex width.4pt depth0pt}
\def\IC{\relax\hbox{$\inbar\kern-.3em{\rm C}$}}
\def\IR{\relax{\rm I\kern-.18em R}}
\def\IN{\relax{\rm I\kern-.18em N}}
\def\IZ{{{\rm Z}\!\!{\rm Z}}}

\def\Tr{{\rm Tr}}

\def\e{~{\rm e}}
\makeatletter
\newdimen\normalarrayskip              
\newdimen\minarrayskip                 
\normalarrayskip\baselineskip
\minarrayskip\jot
\newif\ifold             \oldtrue            \def\new{\oldfalse}
\def\arraymode{\ifold\relax\else\displaystyle\fi} 
\def\@arrayskip{\ifold\baselineskip\z@\lineskip\z@
     \else
     \baselineskip\minarrayskip\lineskip2\minarrayskip\fi}
\def\@arrayclassz{\ifcase \@lastchclass \@acolampacol \or
\@ampacol \or \or \or \@addamp \or
   \@acolampacol \or \@firstampfalse \@acol \fi
\edef\@preamble{\@preamble
  \ifcase \@chnum
     \hfil$\relax\arraymode\@sharp$\hfil
     \or $\relax\arraymode\@sharp$\hfil
     \or \hfil$\relax\arraymode\@sharp$\fi}}
\def\@array[#1]#2{\setbox\@arstrutbox=\hbox{\vrule
     height\arraystretch \ht\strutbox
     depth\arraystretch \dp\strutbox
     width\z@}\@mkpream{#2}\edef\@preamble{\halign \noexpand\@halignto
\bgroup \tabskip\z@ \@arstrut \@preamble \tabskip\z@ \cr}%
\let\@startpbox\@@startpbox \let\@endpbox\@@endpbox
  \if #1t\vtop \else \if#1b\vbox \else \vcenter \fi\fi
  \bgroup \let\par\relax
  \let\@sharp##\let\protect\relax
  \@arrayskip\@preamble}

\begin{document}
\thispagestyle{empty}
 
\vskip2cm
\begin{center}
{\huge \bf Theta Sectors and Thermodynamics of a Classical Adjoint Gas}\\
\vskip1cm
{\bf  S. Jaimungal \footnote {Supported in part by the Natural Science and Engineering 
Research Council of Canada.}  and  L.D. Paniak 
\footnote{Supported in part by a University of British Columbia 
Graduate Fellowship.} }\\
\bigskip

{\it Department of Physics and Astronomy,\\
University of British Columbia\\6224 Agricultural Road\\
Vancouver, British Columbia, Canada V6T 1Z1}
\vskip4cm
\begin{abstract}
The effect of topology on the 
thermodynamics of a gas of adjoint representation charges interacting 
via 1+1 dimensional $SU(N)$ gauge fields is investigated.
We demonstrate explicitly the existence of multiple vacua 
parameterized by the discrete superselection variable
$k=1, \cdots,N$.  In the low pressure limit, the $k$
dependence of the adjoint gas equation of state is 
calculated and shown to be non-trivial. 
Conversely, in the limit of high system 
pressure, screening by the adjoint charges results in an equation of state  independent of $k$. 
Additionally, the relation of this model to adjoint QCD at 
finite temperature in two dimensions  
and the limit $N \rightarrow \infty$ are discussed.

\end{abstract}
\end{center}
\newpage
\setcounter{page}1
\section{Introduction}
It has been known for a long time now that a gauge theory coupled
to matter can have non-trivial vacuum structure.  For our purposes 
this statement is taken to mean that in the theory there exists a 
family of isolated (vacuum) sectors of which the physical one must
be chosen by a superselection rule.  The classification 
of such `$\theta$-vacua' can be carried out in all cases by examining 
the topological structure of the effective gauge group which acts  
non-trivially on the matter content of the theory. 
Unfortunately, this classification has little to say about the 
consequences of the choice of vacuum on the physics of the system.
Our purpose here is, in the case of two dimensional $SU(N)$ gauge fields 
coupled to a gas of classical adjoint charges, to
explicitly determine the thermodynamics of the system
as function of vacuum sector.

The $SU(N)$ non-Abelian Coulomb gas with adjoint charges in two dimensions 
is a useful model for investigating the topological and 
symmetric properties of gauge theories in higher dimensions. 
In the limit of infinite $N$ this model has a first order phase 
transition which can be interpreted as a deconfining one 
separating a phase with tightly bound fundamental degrees of 
freedom and a plasma-like phase.
This behaviour is analogous to the situation in higher dimensional 
Yang-Mills theory and adjoint QCD where the transition is characterized
by a breaking of the center symmetry of the gauge group and the 
Polyakov loop serves an effective order parameter.
Unfortunately, for the case of finite $N$ which we will be studying here
there is no such phase transition in two dimensions.  However at
finite $N$ the vacuum structure  due to the topological structure of 
the theory is clearly apparent. Moreover the model is explicitly solvable
in the limits of high and low particle density and in these limits 
we will investigate the thermodynamics of the adjoint gas in each sector
of the theory.

The standard method of classifying the multiplicity of vacua
in a particular gauge theory with adjoint matter depends on
identifying the effective gauge group. Here, since gauge
transformations operate by adjoint action on all fields, a transformation
which lies in the center $Z$ is trivial.  Consequently
the true gauge group is the quotient of the gauge group and its
center.  This quotient is multiply connected, as can be seen from the 
relation,
\beq
\pi_1( G/Z)~=~ \pi_0(Z)~=~ Z
\eeq
where $G$ is any semi-simple gauge group. Thus, $Z$ gives a 
classification of gauge fields which are constrained to be flat connections
at infinity. In that case
\beq
\lim_{ |x| \rightarrow \infty}~A_\mu(x)~=
ig^{\dagger}(x)\nabla_\mu
g(x)
\eeq
where $g(x)$ is a mapping of the circle at infinity to the gauge group
$G/Z$. Since $G/Z$ is a symmetry of the Hamiltonian,
we expect that all physical states carry a representation
of $\pi_1(G/Z)$. In the case where the center of the group is Abelian
all of its irreducible representations are one dimensional
and further, when $Z \sim \IZ_{n_1} \times \cdots \times
\IZ_{n_j} $,   we are
lead to a classification of all physical states in terms of 
$j$ generators of $Z$, $\{z_1, \cdots, z_j \}$.
 If ${\cal Z }$ is a unitary realization of $Z$ and 
$| \psi >$ is a physical state we have
\beq
{\cal Z} |\psi > = \e^{ i (z_1 + \cdots +z_j)} | \psi >
\eeq 
The equivalence class of all states which have the same transformation 
properties under ${\cal Z}$ form an isolated (vacuum) 
sector of the theory which
is typically called a $\theta$ sector.  This somewhat abstract 
motivation for the existence of multiple sectors in a theory is
in complete agreement with more concrete constructions of  
$\theta$ sectors in $1+1$ dimensional Yang-Mills theory without
matter content \cite{witten,psz1,psz2}.  The objective of the 
current paper is to develop a framework for extending this analysis
to  systems with arbitrary matter content with particular
attention given to the case of
matter in the adjoint representation of the gauge group.

The thermodynamics of classical charges interacting 
via Abelian and non-Abelian electric forces in one spacial 
dimension has been considered previously  \cite{lenard,nambu}.
Additionally the effect of multiple vacua in the $SU(2)$ adjoint
gas has been considered previously \cite{engel}.
We will consider constant pressure ensembles as these authors but 
we will use a different formalism to construct the partition function 
of the system that can be easily extended to configurations other 
than that of the open line. 

We begin in the next section with a short description of our methods
for constructing the model of $1+1$ dimensional non-Abelian Coulomb gas.
Restricting ourselves to the case of the adjoint charges, we proceed 
with an analysis of the low density/pressure limit of the model. Here
using group theoretic techniques the explicit dependence of the 
equation of state of the adjoint gas on the vacuum parameter $k$ is 
established. Converting to the Fourier domain we find the  
high density/pressure limit of the model is equivalent to solving a 
system of coupled quantum oscillators.  In this limit the equation 
of state is shown to be independent of $k$.  We conclude with 
a discussion of the limit of large rank groups, $N \rightarrow \infty$,
and the connection of our model to adjoint two dimensional QCD.

\section{The Classical Non-Abelian Coulomb Gas}
\setcounter{equation}{0}

\begin{figure}
\begin{center}
\epsfbox[55 445 492 506]{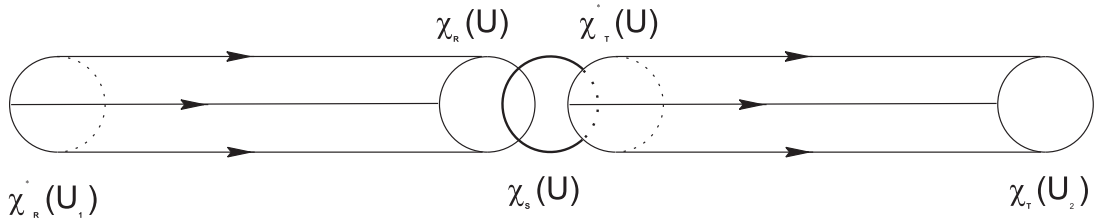}
\caption{{\it To insert matter into the theory the sewing property 
is frustrated by the introduction of a character of representation $S$
at the sewing boundary. In the case of $S$ being the 
trivial representation one recovers (2.3).\label{sewing}}}
\end{center}
\end{figure}

The physical system which we will be investigating is that of a 
collection of static charges restricted to lie on a line, interacting
with each other via 1+1 dimensional non-Abelian electric forces.
Since we are interested in the thermodynamics of this system we will consider
time to be a compactified coordinate with period equal to the 
inverse temperature $\beta = 1/T$.  
Consequently the topology of interest is that of a cylinder.
In this section we demonstrate the construction of a $1+1$-D   
Coulomb gas of non-Abelian charges using
group theoretic ideas introduced by Migdal \cite{migdal} and 
developed later by Rusakov \cite{rusakov}. From this point of view
the propagator for colour electric charge along a spacial distance $L$
with 
boundary holonomies given by the unitary group elements $U_1$ and $U_2$ is,
\beq
K[U_1,U_2;A] = \sum_R \chi_R^*(U_1) ~\e^{-A~C_2(R)}~\chi_R(U_2)
\label{prop1}
\eeq
Here $\chi_R(U)$ is the group character of the element 
$U$ in representation $R$, $C_2(R)$ is the eigenvalue of the
quadratic Casimir operator for the 
representation $R$, and $A\equiv g^2 \beta L/4$ where $g^2$ is 
the gauge field coupling constant.
The characters form an orthonormal basis on the vector space of 
irreducible representations of the gauge group
and hence the propagator (\ref{prop1}) is 
seen to be a diagonal operator on this space.
In the following we will consider the propagator as an operator,
\beq
K[A] = \e^{-A~C_2}
\label{prop2}
\eeq
\begin{figure} 
\begin{center}
\epsfbox[40 422 387 479]{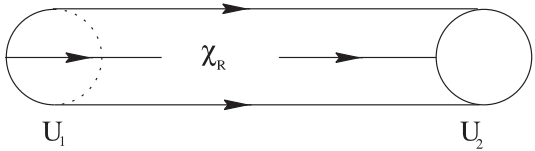}
\caption{{\it The propagator between two open ends on the cylinder 
transports irreducible representations with quadratic Casimir dependent 
exponential damping. \label{prop}}}
\end{center}
\end{figure}
\noindent The convolution of two propagators to form a single propagator
follows from the sewing property \cite{migdal} 
(see fig.\ref{sewing}),
\beq
K[A_1]K[A_2] = K[A_1+A_2]
\eeq 
This relation 
follows directly from the orthogonality of group characters, which
has the effect of multiplying the exponents of (\ref{prop2}) in the naive way.
This sewing is fundamental to using cylinders to build models of 
spherical or toroidal topology and to include external charges.

The inclusion of a static charge of a particular 
irreducible representation $S$ into the system \cite{gpss}
is accomplished by including character operator $\chi_S$ 
at a sewing site (see fig.\ref{sewing}). Here
we will restrict ourselves to the case where all charges in the system
are of the same representation but the extension for arbitrary configurations
is straightforward.
 
Introducing $n$, $S$ representation loops each separated by a distance
$L_1 \cdots L_n$, the operator of interest is
\beq
\prod_{i=1}^n  \left[ \e^{-g^2 \beta L_i C_2/4} \chi_S \right]
\eeq
If one identifies the ends of the cylinder and integrates over each
$L_i$ with an appropriate restriction, we arrive at an expression for 
the partition function of $n$ charges on a circle with total circumference
$L= \sum L_i$.
\beq
Z_n = \frac 1 {n!}  \int_0^L dL_i\dots \int_0^L dL_N ~ \delta
\left ( L- \sum_{i=1}^n L_i \right) 
\Tr \prod_{i=1}^n \left[  \e^{-g^2 \beta L_i C_2 /4} \chi_S \right]
\eeq
\begin{figure} 
\begin{center}
\epsfxsize=5in
\epsfbox[-180 315 562 516]{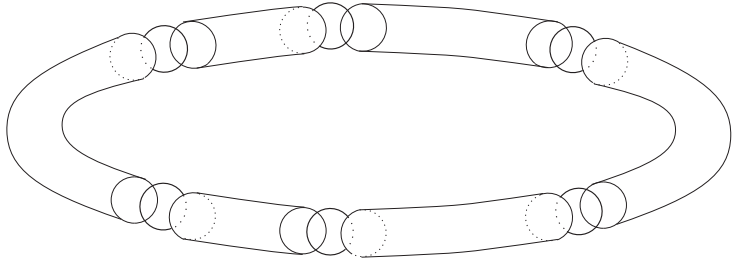}
\caption{{\it Sewing $n$ cylinders together with frustration
and identifying the ends, leaves
one with a torus corresponding to $n$ charges on a circle.}}
\end{center}
\end{figure}
It is worth noting here that this last expression also has the 
interpretation as a gas of Wilson loops on a space-time torus and
with slight modification can be used as a starting point for 
investigating a gas of Wilson loops on the sphere.
Taking the circumference of the circle, $L$ to infinity
effectively reduces the configuration space to an open line. 
Carrying out the now unrestricted
integrations over $\{ L_i\}$ we find that the partition function
takes on a rather simple transfer matrix form
\beq
Z_n =  \frac{1}{n!} \Tr \prod_{i=1}^n \left[
\frac{1}{g^2 \beta C_2/4} \chi_S \right]
\equiv \frac{1}{n!} \Tr T^n
\eeq 
Hence, the calculation of the thermodynamics of 
$n$ static charges on the open line reduces to solving the 
eigenvalue problem for the operator $T$
\beq
T \Psi = \frac{1}{g^2 \beta C_2/4} \chi_S \Psi = \lambda \Psi
\label{evalprobx}
\eeq

For the purpose of finding thermodynamic quantities it is convenient 
to deal not with the constant volume ensemble as we have up 
to now, but rather a constant pressure ensemble.
The change to a constant pressure ensemble
can be carried out in a straightforward manner by introducing
a $p V$ term into the energy of the system with the result of shifting
the energy per unit length.  The resulting eigenvalue problem reads
\beq
T_p \Psi = \frac{1}{\beta ( g^2 C_2/4 +p)} \chi_S \Psi = \lambda \Psi
\label{evalprob}
\eeq
In the  thermodynamical limit where
$n \rightarrow \infty$ all information of the system is contained
in the largest eigenvalue $\lambda_0$ of the operator $T_p$.  The 
remainder of this paper will involve finding $\lambda_0$ for the 
case of a gas of adjoint charges.

\section{Low Density Limit: Group Theory}
\setcounter{equation}{0}

The effective eigenvalue problem for the non-Abelian gas (\ref{evalprob})
was previously derived via a different approach by Nambu et al \cite{nambu}.
As in that case (\ref{evalprob}) is equivalent to 
the following linear equation,
\beq
H \Psi \equiv (\alpha C_2 - q \chi_S   ) \Psi = - p \Psi
\label{tmatrix}
\eeq
which acts on the space of irreducible representations, so that $\Psi=\sum_R
a_R \chi_R$; with $\alpha = g^2/4$ and $q=1/(\beta \lambda)$.
The structure of this equation is the same as one would find 
in a quantum mechanics problem.  The quadratic Casimir is a diagonal 
operator on the space of irreducible representations 
\beq
C_2~ \chi_R = C_2(R)~\chi_R
\eeq
which corresponds to the kinetic term. The role of the potential is 
played by the character $\chi_S$ which mixes the eigenvectors of the 
kinetic term. This can be easily seen by the multiplication rule
\beq
\chi_R ~\chi_S = \chi_{R\otimes S} = \sum_{T} N_{R S}^T \chi_T
\label{fusion}
\eeq
Here $N_{R S}^T$ is the fusion number which enumerates the 
occurrence of the irreducible representation $T$ in the 
Kronecker product of representations $R$ and $S$. 
The only difference between quantum mechanics and the
current situation is that we would like to solve for the eigenvalue 
of the transfer matrix problem $\lambda = 1/(\beta q)$ as a function of the 
pressure $p$ as opposed to solving for the energy of the system as
a function of the potential.

As in the 
case of quantum mechanics one can begin to solve the eigenvalue 
problem by considering the symmetries of the system which will lead to 
conserved quantities. Here we are most interested in the symmetric properties
of the `Hamiltonian', $H$, under transformations which lie in the center
of the gauge group. 
As we have seen, the presence of such a symmetry immediately
leads to the phenomena of multiple vacua. The action of a transformation
under the center of the gauge group is defined as
\beq
{\cal Z} \chi_R = z_R \chi_R
\eeq
Here $z_R$ is a representation of the center of the gauge group. Since
for all compact Lie groups the center forms an Abelian subgroup, we can 
take $z_R$ to be a complex phase factor.  The details of this phase factor 
depend on the structure of the center $Z$.  For $U(N)$, $Z$ is isomorphic to
$U(1)$ hence $z_R = \e^{i a C_1(R)}$ where $a \in \IR$ and
$C_1$ is the integer valued linear Casimir operator. For the case of interest, 
$SU(N)$, $Z\sim \IZ_N$ and consequently $z_R = \e^{2 \pi i C_1(R)/N}$.
This follows from the $U(N)$ case with the restriction that $z_R^N=1$.
The question of whether $H$ commutes with ${\cal Z}$ is reduced, since 
Casimir operators commute amongst themselves, to the question of 
whether
\beq
[\chi_S, {\cal Z}] = ( 1 - \e^{2 \pi i C_1(S)/N}) \chi_S= 0
\eeq
The solutions to such a condition are clearly $C_1(S) = 0$ (mod $N$).
In terms of irreducible representation $S$ of the matter content of the 
theory this means that the full vacuum degeneracy  
is apparent only when $C_1(S) = 0$ (mod $N$).
The simplest examples of such representations are the trivial 
representation in which case the theory reduces to that of pure
Yang-Mills in $1+1$ dimensions and the case where all matter is in 
the adjoint representation.  These are the cases we will consider 
for the remainder of this paper.
 
In the case of the adjoint gas where the center operator commutes
with (\ref{tmatrix}), in analogy with the conservation of eigenvalues
of commuting operators in quantum mechanics we see that the eigenvalue
of the linear Casimir is conserved (mod $N$) and  
is a good `quantum number' (mod $N$). Consequently for the $SU(N)$
adjoint gas there exists a family
of $N$ distinct solutions to the eigenvalue problem each of which we
will label by $k=0, \cdots, N-1$. For each value of $k$ we have an
isolated sector of the theory complete with a stable vacuum and 
an infinite tower of excited states. These are precisely the discrete
`$\theta$-vacua' of the model.

In the limit where $q \rightarrow 0$, the eigenvalue problem 
in (\ref{tmatrix}) is reduced  to that of the free $1+1$
dimensional Yang-Mills theory and we can easily identify the vacuum 
states of each sector.  In this case the eigenvectors, $\Psi$ of the 
transfer matrix, $T_p$
are simply the irreducible representations of the gauge group.
Since we are interested in the case with adjoint $SU(N)$ charges,
the labeling of vacua introduced above with $k=0, \cdots, N-1$
will be followed although strictly speaking the free theory has a
countably infinite vacuum degeneracy.
We denote by $[k]$ the $k^{th}$ 
vacuum state which is the $k(k-1)/2$ dimensional completely antisymmetric 
fundamental representation of $SU(N)$. Each of these 
fundamental representations
is the lowest lying energy state of each of the $k$ sectors and will 
serve as a starting point for a perturbative calculation of the
eigenvalue problem for small $q$, or equivalently, 
small $p$ in each sector.
 
Having identified the ground states of each sector of the theory all 
we need to calculate, to lowest order in the pressure $p$,
the solution of the eigenvalue problem (\ref{tmatrix}) are
the fusion numbers $N_{R S}^T$ (\ref{fusion}).  As we have seen, 
these are pure group theoretic quantities which detail the mixing effect 
of the potential on irreducible representations and, in particular, 
the anti-symmetric ground states $[k]$. 
With this information one can then proceed as in quantum mechanics and 
calculate with the unperturbed ground states the pressure $p$ 
up to third order in  $q$.
Leaving the details of the evaluation of the fusion numbers and 
the perturbative expansion of the eigenvalue problem to the Appendix, we
record the results for the $SU(N)$ adjoint gas in the 
$k^{th}$ sector in Table 1.  
\begin{table}
\begin{center}

\begin{tabular}{|c|l|} \hline

\begin{tabular}{l} $N >2$\rule{0in}{3ex}\\[1ex]$k=0$ \end{tabular} &\rule{0in}{.2in}
\begin{tabular}{l}	
	$\tilde{p}= \tilde{q}^2+\tilde{q}^3 +O(\tilde{q}^4)$ \\[1ex]
	$\rho=\frac{1}{2} - \sqrt{ \frac{\tilde{p}}{2}} + O(\tilde{p}^2)$
\end{tabular}  \rule{0in}{.1in}\\[.2in] \hline

\begin{tabular}{l} $N=3$\rule{0in}{3ex}\\[1ex]$k=1$ \end{tabular} &\rule{0in}{.2in}
\begin{tabular}{l}
	$\tilde{p}= \tilde{q} + \frac{ 9 }{4 }\tilde{q}^2+ \frac{45}{16} \tilde{q}^3 +O(\tilde{q}^4)$\\[1ex]
	$\rho = 1 - \frac{9}{4} \tilde{p} + \frac{153}{16} \tilde{p}^2+O(\tilde{p}^3)$
\end{tabular} \rule{0in}{.1in}\\[.2in] \hline

\begin{tabular}{l} $N > 3$\rule{0in}{3ex}\\[1ex]$k=1$ \end{tabular} &\rule{0in}{.2in}
\begin{tabular}{l}
	$\tilde{p} = \tilde{q} + 2\frac{ N^2 }{N^2 -1 } \tilde{q}^2+ 4\frac{N^4}{(N^2 -1)^2}\tilde{q}^3
	+O(\tilde{q}^4)$ \\[1ex]
	$\rho =1-2\frac{N^2 }{N^2 -1}\tilde{p}+4\frac{N^4}{(N^2 -1)^2}\tilde{p}^2+O(\tilde{p}^3)$
\end{tabular} \rule{0in}{.1in}\\[.2in] \hline

\begin{tabular}{l} $N> 3$\rule{0in}{3ex}\\[1ex]$k=2$ \end{tabular} &\rule{0in}{.2in}
\begin{tabular}{l}
	$\tilde{p}= \tilde{q} + (\frac{N }{2 } + \frac{N}{N+1 }+ \frac{N}{N-2 }) \tilde{q}^2+
		( \frac{ N^3 }{(N -2 )(N+1)}+\frac{N^2}{(N -2 )^2}  
			+ \frac{2N^2}{(N +1 )^2} +)\tilde{q}^3+O(\tilde{q}^4)$\\[1ex]
	$\rho = 1 - (\frac{N }{2 } + \frac{N }{N+1 }+ \frac{N }{N-2 } )\tilde{p}
		+ \frac{3 N^4 + 10 N^3 -13 N^2 - 8 N -24}
		{4(N-2)^2(N+1)^2} N^2\tilde{p}^2 +O(\tilde{p}^3)$
\end{tabular} \rule{0in}{.1in}\\[.2in] \hline

\begin{tabular}{l} $N> 3$\rule{0in}{3ex}\\[1ex]$k>2$ \end{tabular} &\rule{0in}{.2in}
\begin{tabular}{l}
	$\tilde{p}= \tilde{q}+ (\frac{N}{k } + \frac{N }{N+1 }+ 
		\frac{N }{N-k } )\tilde{q}^2 +
		( \frac{N^2}{(N -k )^2}+\frac{N^2}{k^2}+\frac{2N^2}{(N +1 )^2}
		+  
\frac{2 N^3 }{k (N -k )(N+1)} )\tilde{q}^3+O(\tilde{q}^4)\!\!\!\!\!\!$\\[1ex]
	$\rho = 1 - (\frac{N }{k } + \frac{N }{N+1 }+ \frac{N }{N-k } ) \tilde{p}
		+ (\frac{N^2}{(N-k)^2}- \frac{N^2}{(N+1)^2} 
		+ \frac{N^2}{k^2}\frac{5k + N + 7 N k + N^2}{(N-k)(N+1)})\tilde{p}^2
		+O(\tilde{p}^3)\!\!\!\!\!$
\end{tabular} \rule{0in}{.1in}\\[.2in] \hline
\end{tabular}
\caption{{\it Table of the pressure $p$ and the ratio of macroscopic to microscopic
degrees of freedom $\rho$ for the various vacuum states, $k$ of the 
$SU(N)$ adjoint gas. The primed variables are defined
as $\tilde{x}\equiv \frac x{\alpha N}$. Note the conjugation symmetry
$k \rightarrow N-k$. }
\label{density}}
\end{center}
\end{table}

Now we would like to develop the equation of state for system.
As is familiar from more physical gauge theories, the number of microscopic
degrees of freedom $n$ may not be the number of macroscopic  
degrees of freedom, $n^*$. For example it is believed that in QCD
pairs and triples of quarks are bound into observable mesons and bayrons, 
respectively. The equation of state per microscopic degree of freedom
for the constant pressure ensemble is
\beq
\frac{p <V>}{n} \equiv p <v> = \frac{n^*}{n} T
\label{eqst1}
\eeq
where $<V>$ is the expectation value of the total volume of the 
system which is canonically conjugate to $p$.  It can be determined 
by inverting the relation $p(q)$ and using the relationships between 
the thermodynamic variables
\beq
<V> = -T \frac{\partial \log{Z}}{\partial p} =
-n T \frac{\partial \log{\lambda}}{\partial p} = 
n T \frac{\partial \log{q(p)}}{\partial p}
\label{bigv}
\eeq
It is convenient to define $\rho$ as 
the ratio of macroscopic to microscopic degrees of freedom  
\beq
\rho = \frac{n^*}{n} = \frac{\partial \log{q}}{\partial \log{p}}
\label{rhodef}
\eeq
Comparing with (\ref{eqst1}) we see the fundamental importance of 
the quantity $\rho$.
The dependence of $\rho$ on $p$ is tabulated for the different sectors
of the $SU(N)$ gas in Table 1. 

The results of these calculations deserve some comment. The most striking
between the different sectors of the theory is the configuration of 
adjoint charges in the limit of vanishing pressure.  For the $k=0$
sector we find that $\rho=1/2 - \cdots$ and hence the adjoint 
charges in the system are bound pairwise in the low pressure limit.
This behaviour is not surprising and is seen in both the $U(1)$ \cite{lenard}
and $SU(N)$ \cite{nambu} one dimensional (fundamental representation)
Coulomb gases.
What is different here in the $k=0$ sector is that the first 
corrections in pressure to this pair-wise binding come about 
with a negative sign and so the adjoint charges begin to form 
macroscopic configurations where the number of constituents is 
three of more. This is possible since adjoint charges are of course 
self-adjoint and an arbitrary number of them can form an
observable charge singlet.

When one moves to the cases when $k>0$ we see a distinct change in 
the vanishing pressure macroscopic structure of the theory.
As explained previously for the Yang-Mills case \cite{witten,psz1},
different sectors of the a $1+1$ dimensional gauge theory are
equivalent to considering a the theory with different
constant background colour electric fields.  For $SU(N)$, 
each admissible background is given by one of the $N$ fundamental 
representations which we label by the parameter $k$. In Table 1 we
see that for a non-trivial background $(k>0)$ the adjoint charges of 
the system can interact with the background electric field and 
form stable, colour singlet 
configurations where they are the macroscopic degrees 
of freedom. In other words they act as free particles.

\section{High Density Limit: Fermions on the Circle}
\setcounter{equation}{0}

The eigenvalue problem of (\ref{tmatrix}) can also be solved exactly 
in the limit of large values of $q$ which corresponds to the limit
of high pressure.  This is most conveniently carried out by converting
the group theoretic equation of (\ref{tmatrix}) to a linear differential 
equation with periodic coefficients. In this section the gauge group
will be taken to be $U(N)$ as this will simplify 
calculations. Recovering the results for $SU(N)$ is a trivial 
step which will be noted at the appropriate point in the calculation.

The starting point for converting the eigenvalue problem of (\ref{tmatrix})
to a differential equation is to consider the eigenvector $\Psi$ as
a linear combination of irreducible representations each labeled by 
$N$ integers $\{ n_1, \cdots, n_N \}$.  These integers correspond to 
reduced row variables for the Young diagram associated with the 
irreducible representation, and satisfy the dominance condition
\beq
\infty >n_1 >n_2> \cdots>n_N>-\infty
\label{domcond}
\eeq
The quadratic Casimir operator is diagonal in this basis with the action
\beq
C_2 \psi_{n_1, \cdots,n_N} = \frac{1}{2} \left( \sum_{i=1}^{N} n_i^2 - \frac{N(N^2-1)}{12} \right) \psi_{n_1, \cdots,n_N}
\eeq
The action of the character $\chi_S$ on these states is, however, more
complicated. This is partly due to the dominance restriction on
the Young diagram. If one relaxes this restriction then the action of the 
character is simplified somewhat. In this case the action 
on a general state of a character in the adjoint representation 
is given by,
\beq
\chi_{Ad} \psi_{n_1, \cdots,n_N} = \sum_{r,s=1}^{N} 
\psi_{n_1+\delta_{r,1}-\delta_{s,1}, \cdots,n_N+\delta_{r,N}-\delta_{s,N}}
\eeq
The result of this operation is to add unity to $n_r$ and then 
subtract unity from $n_s$ and sum over all $r$ and $s$.
Consequently the eigenvalue problem of (\ref{tmatrix}) is 
a difficult recurrence type equation. This type of 
equation is most successfully dealt with by introducing the 
function $~\tilde{\psi}(x_1, \cdots, x_N)$, which is completely 
symmetric under the exchange of arguments, via a
Fourier transform  
\beq
\psi_{n_1, \cdots,n_N} = \int_{0}^{2 \pi} dx_1 \cdots \int_{0}^{2 \pi}
dx_N 
\e^{i \sum_i n_i x_i} \tilde{\psi}(x_1, \cdots, x_N) 
\Delta(x_1, \cdots, x_N)
\label{fourier}
\eeq
The factor $\Delta(\{ x_i \})$ is a Vandermonde determinant defined
as $\prod_{i<j} (x_i -x_j)$ and as such is completely antisymmetric under
the exchange of any two $x_i$'s. This factor is included to force the 
integration to vanish identically for $n_i =n_j$, when $i \neq j$. 
In this way we can effectively impose the 
dominance condition (\ref{domcond}) of the reduced Young diagram
variables in the Fourier domain.  

Acting with the Casimir and the adjoint character on 
(\ref{fourier}) we find the transfer matrix problem of 
(\ref{tmatrix}) is equivalent to the second order linear partial 
differential equation with  periodic coefficients
\beq
\left[ -\frac{1}{2} \sum_{i=1}^{N} \frac{\partial^2}{\partial x_i^2} -
\frac{N}{24} (N^2 -1) - \frac{q}{\alpha}\left [ N + 2 \sum_{i<j} 
\cos{(x_i - x_j)}\right ] \right]\Delta \tilde{\psi} = -\frac p\alpha \Delta
\tilde{\psi}
\label{diffeq}
\eeq
In this form some of the features of the adjoint non-Abelian Coulomb
gas are more apparent.  For example, the center operator ${\cal Z}$ has
a simple interpretation in the Fourier domain 
\beq
{\cal Z} = \e^{i a C_1} = \exp{\left[ i a \sum_{k=1}^N  n_k \right]}=
\exp{\left[  a \sum_{k=1}^N  \frac{\partial}{\partial x_k} \right]}
\label{bigZ}
\eeq
This is exactly the same structure as the translation operator
in quantum mechanics.  Here ${\cal Z}$  generates uniform 
shifts of the coordinates $\{ x_i \} \rightarrow \{ x_i +a \}$.
It is easy to verify that (\ref{diffeq}) has this symmetry and 
from the arguments of the previous section we expect a
continuum of vacua for 
the adjoint gas with $U(N)$ gauge group.  Additionally with  
periodic coefficients and the restriction to completely anti-symmetrized
wavefunctions, $\Delta \tilde{\psi}$, the eigenvalue problem is 
equivalent to that of non-relativistic fermions in a periodic potential.  
This correspondence is familiar from 
matrix models \cite{bipz} and is exploited in the solution
of the large $N$ non-Abelian Coulomb gas \cite{stz,sz,gps}.

In the limit of high densities, or equivalently high pressure, the wave
functions are localized about the minima of the potential. Expanding 
the potential about the local minimum at $\{x_i- x_j =0 \}$
leads to the coupled harmonic oscillator problem
\beq
\left\{ -\frac{1}{2} \sum_{i=1}^{N} \frac{\partial^2}{\partial x_i^2} -
\frac{N}{24} (N^2 -1) +\frac{q}{\alpha}\sum_{i<j}( x_i -x_j)^2
- \frac{N^2 q}{\alpha}   \right\}  \Delta \tilde{\psi} = -\frac p\alpha
\Delta\tilde{\psi}
\eeq
Performing the change of variables to the orthonormal basis $\{u_i \}$ 
given by  
\bea
u_n &=& \frac {\left( (N-n)~x_n - ( x_{n+1} + x_{n+2} + 
	\dots+ x_N ) \right)}{\sqrt{(N-n)(N-n+1)}}\qquad n=1\dots N-1 
\label{diagvars} \\
u_N &=& \frac 1 {\sqrt{N}}\left(x_1+x_2+\dots+x_N\right) \nonumber
\eea
diagonalizes the problem.  In this basis the decoupled oscillator problem is
\beq
\left [ -\frac 1 2 \sum_{i=1}^N \frac {\partial^2}{\partial u_i^2}  + \frac 
1 2 \left ( \frac{2Nq}\alpha \right )\sum_{i=1}^{N-1} u_i^2 \right ] \psi = E 
\psi
\eeq
where,
\beq
E = \frac N{24}(N^2-1)- \frac {p-N^2q}{\alpha}
\eeq
At this point we note that there is no potential for the $u_N$ coordinate
which is to be expected as it describes a center of mass coordinate in the
change of variables (\ref{diagvars}) and the original potential in
(\ref{diffeq}) depends only on relative not absolute 
positions. Consequently the $u_N$
dependence of the problem is only through a phase.  In the case
of $SU(N)$ this center of mass coordinate is restricted but otherwise
behaves exactly as in $U(N)$, entering only as a phase. Regardless of 
the details of this phase we will see it does not contribute to the 
high pressure equation of state of the adjoint gas.
The other modes corresponding to the coordinates $\{u_1 \cdots u_{N-1} \}$
have degenerate frequencies which are easily read off 
the diagonal form
\beq
\omega_N = \sqrt{\frac {2Nq}{\alpha}}
\eeq
Knowing the normal modes of the eigenvalue problem we are now in a position
to find the ground state solution of (\ref{diffeq})  which will 
correspond to the dominant eigenvalue of the transfer matrix problem.
A solution which satisfies the requirement of 
antisymmetry with respect to permutation of the coordinates is given by,
\beq
\Delta \tilde{\psi}
\sim   \prod_{i<j} (x_i - x_j) ~ \exp{\left\{i \frac{M}{\sqrt{N}} 
\sum_i x_i \right\} } \exp \left\{ - \frac 1 2 
\frac{\omega_N}{N} \sum_{i<j} (x_i-x_j)^2 \right \}
\eeq
The parameter $M$ is an integer associated with the center of mass
coordinate and contributes a constant to the energy eigenvalue. Notice that the
potential exponentiated, this is a direct consequence of the normal modes
having degenerate energy. The above state has $N^2-1$ quanta of energy,
\beq
E = (N^2-1) \frac{\omega_N}{2} + \frac 1 2 M^2
\eeq
so that the pressure is given, up to an irrelevant constant by,
\beq
\frac p\alpha = \frac{N^2 q}\alpha - (N^2-1)\sqrt{\frac{Nq}{2\alpha}}
\eeq
Inverting this relation to find $q(p)$ and using the definition 
(\ref{rhodef}) we find, to leading order, the ratio of macroscopic to 
microscopic degrees of freedom
\beq
\rho = 1 - \frac{(N^2-1)}{2} \sqrt{\frac{\alpha}{2Np}} + O(\frac{1}{p})
\label{largePdensity}
\eeq
Consequently, in this high pressure limit the adjoint charges are the 
macroscopic degrees of freedom and, as in the $k>0$ low pressure
cases, can be interpreted as being free.  The most striking feature
of  (\ref{largePdensity}) is the absence of any $k$ dependence.
This follows from the fact that the information of the 
center of mass coordinate appears only as a phase contributing the 
additive constant $M^2$ to the eigenvalue problem which is negligible in 
the limit of large pressure. In terms of physics 
the high pressure adjoint gas effectively screens all colour electric 
fields over large distances and so the fundamental colour electric fields
associated with the different vacuum sectors are washed out by the 
adjoint degrees of freedom.
 
\section{Discussion and Conclusions}
\setcounter{equation}{0}

For any finite $N$ the $SU(N)$ non-Abelian Coulomb gas 
does not admit phase transitions, however, in
the limit of infinite $N$ it has been shown \cite{stz,sz,gps} that a 
phase transition develops for the trivial vacuum sector. In those works,
it was found that the adjoint
`quarks' remain bound strictly pairwise, to form `mesons', up to some 
critical pressure at which a first order phase transition occurs and the
mesons disassociate into free quarks. If one takes the infinite $N$ limit
\footnote{In this limit the product $\alpha N$ is kept at order unity, and 
we use $\tilde{p}$ to denote the reduced pressure $\frac p{\alpha N}$}
in the trivial vacuum sector $(k=0)$, for small pressures 
the ratio of macroscopic to microscopic degrees of freedom is 
(see table \ref{density}),
\beq
\rho = \frac 12 -\sqrt{\frac{\tilde{p}}{2}}+ O\left(\tilde{p}\right)
\label{largeN}
\eeq
which appears to contradict the prediction of the large $N$ calculations.
Equation (\ref{largeN}) predicts that the quarks are bound 
pairwise only at zero pressure, and can form bound structures with more
components slightly away from zero pressure. 
This discrepancy can be explained by noting that in the infinite $N$
limit all pressures are naturally measured in units of $N^2$.
Consequently the statement that quarks are bound pairwise 
from zero pressure up to some 
critical pressure is misleading. The correct statement is: for
pressures of order $N^2$ but less than the critical pressure, the quarks 
are bound strictly pairwise. In this large $N$ limit nothing can be said about pressures of order unity, where our current computations are valid.
We see then that the infinite $N$ computations of \cite{stz,sz,gps}
are complementary to our own analysis. To state it another way,
our computations cannot say anything about the affect that multiple
(vacuum) sectors in the theory have on the phase transition at large $N$.
However in the limit of infinite $N$ our calculations are perfectly
valid in the low pressure region.
Taking the infinite $N$ limit of the density for $N>3$, $k>2$ one finds 
(see table \ref{density}),
\beq
\rho = 1 - \left( \frac 1 {\theta (1-\theta)} + 1\right) \tilde{p} 
+\left( \frac {1+6\theta-6\theta^2}{(1-\theta)^2\theta^2} - 1 \right) 
\tilde{p}^2 + O\left(\tilde{p}^3\right)
\eeq
where $k \equiv \theta N$. This demonstrates the explicit dependence on
vacuum sector in the small pressure region. 

In the high pressure limit the picture is much simpler. The infinite
$N$ limit of the exact expression for the ratio of macroscopic to 
microscopic degrees of freedom $\rho$ is given by
\beq
\rho = 4\hat{p} ~\frac { 1 + (1+8\hat{p})^{-\frac 1 2}}{1+4\hat{p}+
(1+8\hat{p})^{\frac 12}}
\eeq
Here we have taken the pressure to be of order $N^2$ by setting 
$\tilde{p} = \hat{p}~N^2$ before taking the infinite $N$ limit. This ensures
that the expression for the density is in the correct pressure scale. Taking 
the pressure to be of order $N^2$ allows us to probe near the phase transition
region, but not the transition itself, since our high pressure
approximation of the last section has neglected the topology 
of the problem which drives this transition. 
For instance we find that at a pressure
of $\hat{p}=\frac{3}{8}$ the density of the gas is exactly one half. This
gives a lower bound on $\hat{p}_{crit.}$ since the phase transition must occur
before the gas confines. Also at very large pressures we have the 
prediction
\beq
\rho = 1 - \frac{1}{\sqrt{8 \hat{p}}} + O \left( \frac{1}{\hat{p}}\right)
\eeq
which agrees with the large $N$ computations.

Finally, some comments on the application of our results to the 
case of two dimensional $U(N)$ Yang-Mills theory coupled to 
adjoint matter.  For adjoint 
representation  fields with mass $m$ the high
temperature effective potential can be calculated \cite{weiss,gpy,kut}
\beq
V(x_i - x_j) \sim \sum_{k=1}^{\infty} \frac{(-1)^k}{k} K_1(m \beta k) 
\cos{[ k (x_i -x_j)]}
\label{kutdet}
\eeq
Here $K_1$ is a modified Bessel function.  In the limit of large
mass such that $m \beta$ is large it can be shown
that the effective potential (\ref{kutdet}) reduces to the form of
the transfer matrix problem (\ref{diffeq}) we have considered previously.
For the case of finite mass, the full high temperature potential is 
difficult to deal with but it clearly has a symmetry under 
translations of the form $\{x_i \} \rightarrow \{ x_i +a \}$.
As we have seen previously, such a symmetry is synonymous with invariance
of the model under center transformations (\ref{bigZ}).
This strongly suggests that our results for the classical adjoint
gas carry forward to massive adjoint two dimensional QCD at high 
temperature and we should expect multiple vacua to exist in that
theory.  Moreover, the physics in each sector of the theory will
depend on the superselection parameter which labels the sectors.

In conclusion,
from consideration of the thermodynamics of a system of static 
adjoint representation charges interacting via $SU(N)$ colour
electric fields in $1+1$ dimension we have shown that the
physics depends on the discrete vacuum index $k$. We have solved the model 
in two regions: low and high pressure. In the low pressure regime,
which is equivalent to low particle density, the constant pressure 
equation of state was shown to have strong dependence on $k$. 
In the limit of high pressures/particle
densities the dependence on $k$ was shown to become trivial, and does
not enter into the equation of state for the adjoint gas.
This is attributed to the screening nature of the high pressure 
limit, which washes out any global structure like a 
vacuum index. 
 
\section*{Acknowledgments}
We would like to thank G. Semenoff, A. Zhitnitsky  
for helpful discussions. Also we would like to thank the 
Asia Pacific Center for Theoretical Physics (APCTP) and the 
Pacific Institute of Mathematical Sciences (PIms) for hosting 
the Summer Workshop on Field Theory, Strings and Mathematical Physics
during which part of this work was completed.
 
\section*{Appendix}
\renewcommand{\thesection}{A}
\setcounter{equation}{0}

Here we present the details of the perturbative calculation of the 
lowest lying eigenvalue of equation (\ref{tmatrix})for the adjoint gas.  
This can  be accomplished via the familiar time independent perturbation 
theory of quantum mechanics.  The problem we are faced with is 
\beq
(\alpha C_2 - q \chi_{Ad}   ) \Psi = - p \Psi
\label{appeq}
\eeq
As noted previously the eigenvectors of the Casimir operator $C_2$ are the 
irreducible representations of the gauge group so a convenient basis in the 
limit as $ q \rightarrow 0$ is $\Psi = \sum a_R \chi_R$.  Also we have
seen that the ground state of each sector of the theory is given 
by one of the $N$ fundamental representations, $\{ [k] \}$ of $SU(N)$.  
Consequently our objective is to calculate the perturbations to the 
pressure $p$ for small $q$ for (\ref{appeq}) about each ground
state $[k]$.  In order to carry out this calculation we need to know 
the matrix elements of the potential in the basis of the irreducible 
representations.  This information follows from the Kronecker product of 
the adjoint representation with our chosen basis
\beq
\chi_{Ad}  \chi_R = \chi_{Ad \otimes R}= \sum_{T} N_{Ad ~ R}^T \chi_T
\eeq
Here $N_{Ad ~R}^T$ is the fusion number enumerating the occurrence 
of the irreducible representation $T$ in the product of the adjoint 
representation $Ad$ and $R$.
As in quantum mechanics we can easily calculate the corrections
to $p$ up to third order in $q$ using the unperturbed basis of irreducible 
representations.
\bea
p_{[k]} &=&  q N^{[k]}_{Ad~[k]} + q^2\sum_{R\ne [k]}
\frac{\left(N^{R}_{Ad~[k]}\right)^2}{C_2([k])-C_2(R)} \nonumber\\
&&\qquad+ q^3\Bigg[\sum_{R,S\ne[k]}
\frac{ N^{[k]}_{Ad~S} N^{S}_{Ad~R} N^{R}_{Ad~[k]}}{\left(C_2([k])-C_2(R)
\right)\left(C_2([k])-C_2(S)\right)}  \label{peval}\\
&&\qquad+N^{[k]}_{Ad~[k]} \sum_{R\ne[k]} \left(\frac{N^{R}_{Ad~[k]}}
{C_2([k])-C_2(R)}\right)^2\Bigg] + O(q^4) \nonumber
\eea
It should be noted that we have left out a constant, sector dependent,
background contribution to the pressure.

In order to give the details of the calculation we need a notation to 
label the irreducible representations. The one we will use here is  
given by the column variables $[m_1,m_2,\dots]$ of the Young diagram 
associated with the representation.  For example the antisymmetric combination
of $k$, $N$ dimensional fundamental representations in $SU(N)$ - the 
ground state of the $k^{th}$ sector - 
corresponds to a Young diagram with a 
single column of $k$ boxes: $[k]$.  Another example which appears in 
all calculations is that of the adjoint representation $(Ad)$ which 
in column variables is given by: $[ N-1,1]$.  In this notation the 
quadratic Casimir for a representation $[m_1,m_2,\dots]$ is given 
by
\beq
C_2(R)=V
\pmatrix{1-\frac 1N & 1-\frac 2N & 1-\frac 3N &\dots &1-\frac{N-1}{N}\cr
	 1-\frac 2N & 2(1-\frac 2N) & 2(1-\frac 3N) &\dots& 2(1-\frac{N-1}N)\cr
	 1-\frac 3N & 2(1-\frac 3N) & 3(1-\frac 3N) &\dots& 3(1-\frac{N-1}N)\cr
	\vdots	& \vdots & \vdots & \ddots &\vdots\cr
	1-\frac{N-1}N & 2(1-\frac{N-1}N) & 3(1-\frac {N-1}N) & \dots 
			&(N-1)(1-\frac{N-1}N) \cr}V^T 
- \frac N{12}(N^2-1)
\eeq
where
\bd
V=\left( m_1+1, m_2+1, \dots, m_{N-1}+1\right)
\ed

The remaining task is to compute the relevant fusion numbers. We begin 
by presenting the results of the calculations in Table 2. Here we record 
the fusion numbers $N_{Ad ~R}^S= N_{Ad ~S}^R$ for 
the representations $R$ and $S$ of importance in the calculation 
(\ref{peval}) of the pressure.  Each sub-table corresponds to a 
different background $k$ for $SU(N)$ since the details of the calculation of fusion numbers in general depends on $k$ and $N$.  
These results are only good for $k \leq N/2$ where the remainder of the
cases can be found via the symmetry of the eigenvalue problem under
conjugation $k\rightarrow N-k$.
\begin{table}
\begin{tabular}{|r|cc|} \hline
$k=0,~N\ge 2$&$[0]$ & $[N-1,1]$ \rule{0in}{3ex} \\[1ex] \hline
$[0]$ 	& 0  &  1   \rule{0in}{3ex}\\ 
$[N-1,1]$& 1  &  2  \\ \hline
\end{tabular}

\begin{tabular}{|r|ccc|} \hline
$k=1,~N=3$&$[1]$ & $[2,2]$ & $[2,1,1]$\rule{0in}{3ex} \\[1ex] \hline
$[1]$ 	& 1  &  1    &   1     \rule{0in}{3ex}\\ 
$[2,2]$	& 1  &  1    &   1 \\
$[2,1,1]$& 1  &  1    &   2 \\ \hline
\end{tabular}

\begin{tabular}{|r|ccc|}\hline
$k=1,~N\ge4$&$[1]$&$[N-1,2]$	&$[N-1,1,1]$ \rule{0in}{3ex} \\[1ex] \hline
$[1]$	& 1	& 1		& 1 \rule{0in}{3ex}\\
$[N-1,2]$&1	& 2		& 1 \\
$[N-1,1,1]$&1	&	1	& 2 \\ \hline
\end{tabular}

\begin{tabular}{|r|cccc|}\hline
$k=2,~N\ge3$&$[2]$&$[N-1,3]$&$[1,1]$&$[N-1,2,1]$ \rule{0in}{3ex}\\[1ex]\hline
$[2]$	&1&1&1&1 \rule{0in}{3ex}\\
$[N-1,3]$&1&2&0&1\\
$[1,1]$&1&0&1&1\\
$[N-1,2,1]$&1&1&1&3\\ \hline
\end{tabular}

\begin{tabular}{|r|cccc|}\hline
$k\ge3,~N\ge3$&$[k]$&$[N-1,k+1]$&$[k-1,1]$&$[N-1,k,1]$ \rule{0in}{3ex}\\[1ex]\hline
$[k]$	&1&1&1&1 \rule{0in}{3ex}\\
$[N-1,k+1]$&1&2&0&1\\
$[k-1,1]$&1&0&2&1\\
$[N-1,k,1]$&1&1&1&3\\ \hline
\end{tabular}
 \caption{{\it Table of relevant fusion numbers 
$N_{Ad ~R}^S =N_{Ad ~S}^R$ for the  
calculation of the pressure
of the adjoint gas. Note these results hold only for $k \leq N/2$ with 
the other cases given by the symmetry $k \rightarrow N-k$.} }
\end{table}
\noindent  
For completeness we present the details of the second table
for $k=1$ and $N=3$.  This is the familiar case of $SU(3)$ and
via common tensor or Young diagram methods the Kronecker 
products of the $8$-dimensional 
adjoint representation $(Ad)$ with the lowest lying 
representations can be calculated.  In dimension notation 
we have
\bea
8\otimes 3 &=& 3 \oplus 6 \oplus 15 \nonumber \\
8\otimes 6 &=& 3 \oplus 6 \oplus 15 \oplus  \cdots  \\ 
8\otimes 15 &=& 3 \oplus 6 \oplus 2 \times 15 \oplus  \cdots \nonumber
\eea
In the last two products we have ignored higher representations
which do not contribute to $O(q^3)$ in (\ref{peval}).
Converting to our column notation
\bea
3 & \equiv & [1] \\\nonumber
6 & \equiv & [2,2] \\
8 & \equiv & [2,1] \\ \nonumber
15 & \equiv & [2,1,1] \\ \nonumber
\eea
we have the results of the second sub-table in Table 2.

\end{document}